# An automated method to calculate the arrival time of shear waves for bender element tests

Tarun Naskar,

*Indian Institute of Technology, Madras-600036, India*

**ABSTRACT**

Bender element is used to determine the shear modulus of the soil sample, and to accurately predict the shear modulus of any sample, it required to preciously locate the advent point of the shear wave component in the recorded output signal. All the existing methods enforces to identify those point manually. The lack of automation makes the process tedious, time consuming and prone to error. In addition, contamination from the P wave and the presence of noises makes this identification procedure dubious. It requires an expert judgment to overcome the above mentioned difficulties. In this paper, a new method is suggested based on "Sliding Fourier Transform" to resolves all these complications. The proposed method can be automated easily and almost requires no human inputs. The method found to be reliant against the presence of noise, P wave contamination and distortion of the signal. The proposed method found to be a superior replacement of existing methods.

Corresponding author. Tel.: +91 44 2257 4322;.

Email addresses:  tarunnaskar@iitm.ac.in (T. Naskar)

**INTRODUCTION**

Although the bender element test was introduced for geotechnical testing in the late 1970s (Shirley and Hampton 1978, Shirley 1978), interpretation of its results remains debatable till present day. Despite several recommendation suggested by many researchers (Airey and Mohsin 2013, Alvarado and Coop 2011, Arroyo et al. 2003, Viggiani and Atkinson 1995, Brignoli et al. 1996a,b , Arulnathan et al. 1998 , Blewett et al. 1999, 2000, Greening and Nash 2004, Lee and Santamarina 2005, Styler and Howie 2013, Camacho et al. 2015, Da Fonseca et al. 2009, Wang et al. 2007, Schultheiss 1981, Dyvik and Madshus 1985, Nash et al. 1997, Lings and Greening 2001, Clayton et al. 2004, Rio 2006), accurately determining the arrival time of shear wave component still poses a challenge . Among all these methods, first time arrival and peak to peak method became the most popular among researchers. Yamashita et al. (2009) tried to find a unifying testing procedure for the bender element. They performed numerous tests at 23 different laboratories located across 11 different countries. They concluded that the first time of arrival (start to start) method produced most consistent results compare to other methods. Still, they got contradictory results when data from different laboratories were compared. In such a problematic scenario, they concluded that an experienced professional is required to manually select the arrival time of the shear wave component.

In the present article, a mathematical operation is suggested to solve the uncertainty related to the manual determination of arrival time of shear wave component. The method is based on "sliding transform Fourier transform (SLFT)" proposed by Kumar and Naskar (2017) to solve the ambiguity related to unwrapping of wrapped phase related to the spectral analysis of surface wave (SASW) test data. Originally developed for minimum two sensors, the SLFT method was modified to work for single sensor data typically recorded by bender element tests. In order to validate the effectiveness of this proposed method, a series of bender element was conducted. A particular emphasis has been placed on determining the number of unwrapped cycles. The results obtained were compared with the existing methods using data obtained from the same tests.

**PROPOSED METHOD**

The new method proposed in this paper is based on sliding Fourier transform introduced by Kumar & Naskar (2017) to solve the phase unwrapping problem. Sliding Fourier transform or SLFT is a variation of Fourier transform which was used to calculate the arrival time of Rayleigh wave phase velocity by using at least two sensors data. Unlike short Fourier transform (SFT) it's not a tradeoff between frequency and time resolution. The method is also found to be very robust and can be automated easily. However, for bender element test it is not possible to use SLFT directly to calculate the arrival time of shear wave component. Because, in the bender element test, use of multiple sensors to measure output signal at different length of the same sample is not feasible. To overcome this problem, SLFT is modified in such a way that now it can be applied to calculate the arrival time of a wave from a single sensor data. The method proposed by the authors in this paper can be implemented with four easy steps.

***Step-I, Determining dominant frequency content:*** Let $f(t)$ is the recived signal from a bender element test. Then the dominant frequency contained in the output signal is calculated using a fast Fourier transform (FFT).

$$F(\omega) = \int_0^T f(t) e^{-i\omega t} dt$$

Where $\omega$ is the circular frequency. The converted signal is then passed through a band pass filter to eliminate noises present in the signal (Fig. 1a). The limit for the cutoff frequencies regarding the band pass filters depends on individual cases, but from the vast amount of data analyzed, the author is proposing to a cutoff frequency of 5 kHz for high pass filter for all the output signals with an exciting frequency between 10 kHz to 100 kHz. Let assume, $\omega^* = 2\pi f^*$ is the dominant frequency component in the filtered signal, that is

$$|F(\omega^*)| \geq |F(\omega)| \text{ For all } \omega^* \in \omega$$

Due to its amplitude, the dominant frequency's influence on the output signal will be relatively greater when compared to other frequencies present in the signal. That is why, if a single frequency is used to analyze the output signal's nature, the dominant frequency will be the most suitable one.

***Step-II Applying SLFT:*** The fast Fourier transform, in a nutshell, is nothing but multiplying the signal with a group of harmonics with different frequencies. If a particular frequency is present in the signal, the multiplication will result in segregation of its contribution to the original signal separated from the contribution of other frequencies.

$$\int_{-\pi}^{\pi} \sin nx \sin mx = \begin{cases} \pi, & m = n \\ 0, & m \neq n \end{cases}$$

$$\int_{-\pi}^{\pi} \cos nx \cos mx = \begin{cases} \pi, & m = n \\ 0, & m \neq n \end{cases}$$

The multiplication product will be maximum once the phase of the frequency presented in the signal and phase of the multiplying harmonics with same frequency matches each other. The SLFT utilize this additional property and slide the signal to align the output signals phase with that of multiplying sinusoidal frequency (Fig 1b).

$$\bar{F}(\psi, t) = \int_0^T f(t) \sin(\omega^* t + \psi) dt$$

Where $\psi$ is the phase angle and its maximum and minimum value are kept equal to 0 and $2\pi$ and $F_s$ = sampling rate. Let assume, at $\psi = \psi^*$ the function $\bar{F}(\psi, t)$ becomes maximum, i.e. $\bar{F}(\psi^*, t) \geq \bar{F}(\psi, t)$, for all $\psi^* \in \psi$.

To calculate at which exact time the signal has arrived, one need to examine the product of the harmonic signal and output signal for each cycle of harmonic. This is also known as phase unwrapping of the wrapped signal. Now, any signal like water waves will always arrive from zero at any specific instance of time and it will be having a shape of a sine curve with an exponential decay. So, the time axis of the signal needs to be adjusted using $t_0^* = \frac{\psi^*}{2\pi f^*}$ and arrival cycle can be calculated by:

$$\varphi(\eta) = \int_{(\eta-1) \times \frac{1}{f^*}}^{\eta \times \frac{1}{f^*}} f(t_0^* + t) \sin(\omega^* t) dt$$

Where $\eta$=0, 1, 2, 3……..$\frac{T}{f^*}$. Now in case of time bounded signal at the ideal situation without the presence of any noise the value of $\varphi$ will be zero for all the cycle ($\eta$) of recorded data when the signal was not arrived and non-zero when the signal is present in the record (Fig 2c). But in all practical situation there going to be always some noise plus there going to some significant contribution from P wave which arrives just before S wave. This P wave often masked the arrival point of S wave and creates the ambiguity at first place. The present method can filter out the contribution of P wave component and the arrival point of S wave component will be indicated by significant increase in the value of $\varphi$. The author is proposing cycle ($\eta$) that's is adjacent to the maximum $\varphi$ which crosses 30% of maximum $\varphi$ or more as an indicator of arrival the of S wave.

***Step-IV,*** Let assume at $\eta = \eta^*$ mark the arrival point of S wave, Then total time taken ($T_s$) for S wave of frequency $\omega^*$ is determined by

$$T_s = t_0^* + \frac{1}{f^*} * \eta^*$$

If the length of the sample is assumed to be $l$, then the velocity of the S wave belongs to a frequency $\omega^*$ is:

$$V_s = \frac{l}{T_s}$$

Now, if the property of the material is not changing in the direction of wave propagation, the shear wave velocity will remain constant irrespective of frequency. Thus $V_s$ will represent the overall velocity of S the wave.

*ADVANTAGE OF PRESENT METHOD:*

One of the biggest drawbacks of the existing methods is, their inability to fully automate the procedure. The peak to peak and arrival time method, both required manual selection of the starting point of the shear wave. This makes the procedure extremely time consuming and susceptible to human judgmental error. All the steps required in the present method proposed by the author can be automated easily with basic knowledge of any computational language. Thus making the procedure fast and human error free. The computational cost of the present method found to be nominal. The method is also found to be resilient against the presence of high level of noise.

In some cases, bender element test data were found to be distorted and repeated testing does not improve the situation. In these cases, the recorded data was either translated above the time axis or rotated or can be both. The present method performs well in these cases as it does not require the signal to cross time axis physically. This is also helpful in case the shear wave arrival point located in-between a wave cycle and not at the beginning of the. Another problem in finding shear wave advent time by existing methods is the contamination of the output signal by P wave component. The P wave generally has higher velocity. Thus it arrives just before Shear wave masking the advent point of S wave components. Sometimes it becomes extremely difficult to locate the S wave arrival point with modest accuracy. The present method can separate the P wave component from the output signal. Generally, P wave has lower amplitude compare to S wave, also P wave has different frequency content than S wave. Thus the present method can recognize them using low amplitude cycle before the arrival of S wave. When the S wave component arrives at the output signal, the present method reacts with a sudden jump in the measured amplitude.

**CONCLUSION:**

A new method has been proposed to determine the shear wave velocity of the bender element test sample. Unlike existing methods, which relies on manual identification of the starting point of the S wave on the recorded signal, the proposed method can be automated with easy four steps. The methods were found to be robust, fast and computationally less demanding to be implemented. Also, the method was found to be quite resilient against the presence of noise and difficulties like a distorted signal. The method can separate the influence of P wave fro S wave in the recorded data to a great extent. Overall, it is expected that the proposed method is a better solution compared to all existing method for finding shear wave velocity of blender element test sample.

**REFERENCES**

Airey, D., and Mohsin, A.K.M. 2013. Evaluation of shear wave velocity from bender elements using cross-correlation. Geotechnical Testing Journal, 36(4): 20120125. doi:10.1520/GTJ20120125.

Alvarado, G., and Coop, M.R. 2011. On the performance of bender elements in triaxial tests. Géotechnique, 62(1): 1–17.

Arroyo, M., Muir Wood, D., and Greening, P.D. 2003. Source near-field effects and pulse tests in soil samples. Géotechnique, 53(3): 337–345..

Arulnathan, R., Boulanger, R.W., and Riemer, M.F. 1998. Analysis of bender element tests. Geotechnical Testing Journal, 21(2): 120–131.

Blewett, J., Blewett, I.J., and Woodward, P.K. 1999. Measurement of shear-wave velocity using phase-sensitive detection techniques. Canadian Geotechnical Journal, 36(5): 934–939.

Blewett, J., Blewett, I.J., and Woodward, P.K. 2000. Phase and amplitude responses associated with the measurement of shear-wave velocity in sand by bender elements. Canadian Geotechnical Journal, 37(6): 1348–1357.

Brignoli, E.G.M., Gotti, M., and Stokoe, K.H. 1996a. Measurement of shear waves in laboratory specimens by means of piezoelectric transducers. Geotechnical Testing Journal, 19(4): 384–397.

Brignoli, E.G.M., Gotti, M., and Stokoe, K.H. 1996b. Measurement of shear waves in laboratory specimens by measurement of Piezoelectric Transducers. Geotechnical Testing Journal, 19(4): 384–397.

Camacho-Tauta, J.F., Cascante, G., VIANA DA FONSECA, A., and SANTOS, J.A. 2015. Time and frequency domain evaluation of bender element systems. Géotechnique, 65(7): 548–562.

Clayton, C.R.I., Theron, M., and Best, A.I. 2004. The measurement of vertical shear-wave velocity using side-mounted bender elements in the triaxial apparatus. Géotechnique, 54(7): 495–498. doi:10.1680/geot.2004.54.7.495.

Da Fonseca, A.V., Ferreira, C., and Fahey, M. 2009. A framework interpreting bender element tests, combining time-domain and frequency-domain methods. Geotechnical Testing Journal, 32(2): 91–107.

Dyvik, R., and Madshus, C. 1985. Lab measurements of Gmax using bender elements. In Advances in the art of testing soils under cyclic conditions. pp. 186–196.

Greening, P.D., and Nash, D.F.T. 2004. Frequency domain determination of G0 using bender elements. Geotechnical Testing Journal, 27(3): 288–294.

Kumar, J., and Naskar, T. 2017. Resolving phase wrapping by using sliding transform for generation of dispersion curves. Geophysics, 82(3): 127–136..

Lee, J.-S., and Santamarina, J.C. 2005. Bender elements: performance and signal interpretation. Journal of Geotechnical and Geoenvironmental Engineering, 131(9): 1063–1070.

Lings, M.L., and Greening, P.D. 2001. A novel bender / extender element for soil testing. Géotechnique, 51(8): 713–717.

Nash, D.F.T., Pennington, D.S., and Lings, M.L. 1997. Anisotropy of G0 shear stiffness in Gault clay. Géotechnique, 47(3): 391–398.

Rio, J.F.M.E. 2006. Advances in laboratory geophysics using bender elements. Available from http://discovery.ucl.ac.uk/1348997/.

Schultheiss, P.J. 1981. Simultaneous measurement of P & S wave velocities during conventional laboratory soil testing procedures. Marine Geotechnology, 4(4): 343–367.

Shirley, D.J. 1978. An improved shear wave transducer. The Journal of the Acoustical Society of America, 63(5): 1643–1645. Acoustical Society of America. doi:10.1121/1.381866.

Shirley, D.J., and Hampton, L.D. 1978. Shear-wave measurements in laboratory sediments. The Journal of the Acoustical Society of America, 63(2): 607–613.

Styler, M.A., and Howie, J.A. 2013. Combined time and frequency domain approach to the interpretation of bender-element tests on sand. Geotechnical Testing Journal, 36(5): 649–659.

Viggiani, G., and Atkinson, J.H. 1995. Interpretation of Bender Element Tests. Géotechnique, 45(1): 149–154.

Wang, Y.H., Lo, K.F., Yan, W.M., and Dong, X.B. 2007. Measurement biases in the bender element test. Journal of Geotechnical and Geoenvironmental Engineering, 133(May): 564–574.

Yamashita, S., Kawaguchi, T., Nakata, Y., Mikami, T., Fujiwara, T., and Shibuya, S. 2009. Interpretation of international parallel test on the measurement of Gmax using bender elements. Soils and Foundations, 49(4): 631–650.

**Figure Captions**

Fig. 1. (a) Step-I, Frequency spectrum of received signal and filtering out low frequency noise (b) Step-II, Sliding the output signal to match harmonic multiplying factor (c) Step-III, Unwrapping the wrapped phase .

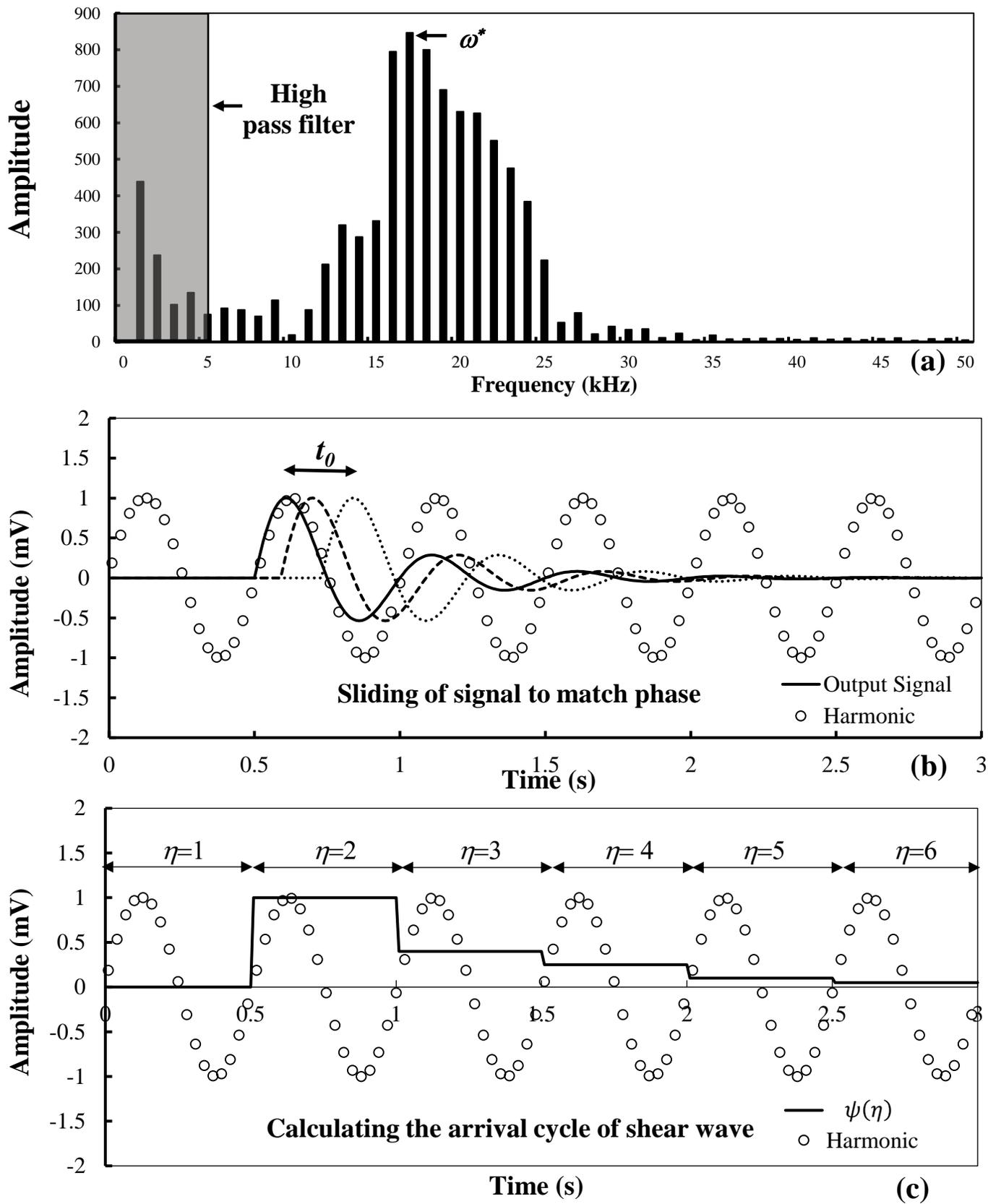

**Fig. 1. (a)** Step-I, Frequency spectrum of received signal and filtering out low frequency noise **(b)** Step-II, Sliding the output signal to match harmonic multiplying factor (c) Step-III, Unwrapping the wrapped phase.